\documentclass[journal]{IEEEtran}

\usepackage{pifont}
\usepackage{eurosym}
\usepackage{amsmath}
\usepackage{amssymb}
\usepackage{mathtools}
\usepackage[utf8]{inputenc}
\usepackage[vietnamese,english]{babel}
\usepackage{cite}
\usepackage{slashbox}

\ifCLASSINFOpdf
\usepackage[pdftex]{graphicx}
\else
\fi

\usepackage{amsmath}

\usepackage{url}

\usepackage{makecell}
\usepackage[subtle]{savetrees}
\usepackage[table, dvipsnames]{xcolor}

\usepackage[acronym,toc,shortcuts]{glossaries}
\usepackage{csquotes}

\ifCLASSOPTIONcompsoc
\usepackage[caption=false,font=normalsize,labelfon
t=sf,textfont=sf]{subfig}
\else
\usepackage[caption=false,font=footnotesize]{subfi
g}
\fi

\newacronym{2D}{2D}{two dimensional}
\newacronym{3D}{3D}{three dimensional}
\newacronym{5G}{5G}{Fifth Generation}
\newacronym{5G PPP}{5G PPP}{5G Infrastructure Public Private Partnership}
\newacronym{3GPP}{3GPP}{3rd Generation Partnership Project}
\newacronym{4D}{4D}{four dimensional}
\newacronym{AAA}{AAA}{Authentication, Authorization and Accounting}
\newacronym{ABS}{ABS}{Aerial Base Station}
\newacronym{ABSF}{ABSF}{Almost-Blank Subframe}
\newacronym{AES}{AES}{Advanced Encryption Standard }
\newacronym{AI}{AI}{Artificial Intelligence}
\newacronym{AMC}{AMC}{Adaptive Modulation and Coding}
\newacronym{AP}{AP}{access point}
\newacronym{API}{API}{Application Programming Interface}
\newacronym{APN}{APN}{Access Point Name}
\newacronym{AWGN}{AWGN}{additive white Gaussian noise}
\newacronym{BBU}{BBU}{Baseband Unit}
\newacronym{BE}{BE}{best-effort}
\newacronym{BET}{BET}{Blind Equal Throughput}
\newacronym{BLAST}{BLAST}{Bell Laboratories Layered Space-Time}
\newacronym{BLER}{BLER}{Block Error Rate}
\newacronym{BS}{BS}{Base Station}
\newacronym{BTP}{BTP}{Backhaul Transport Provider}
\newacronym{BTS}{BTS}{Base Transceiver Station}
\newacronym{CA}{CA}{carrier aggregation}
\newacronym{MBAR}{MBAR}{Mobile Backhaul Aggregation Router}
\newacronym{CAPEX}{CapEx}{capital expenditure}
\newacronym{CDF}{CDF}{Cumulative Distribution Function}
\newacronym{CELL-ID}{CELL-ID}{cell identification ID}
\newacronym{CIO}{CIO}{cell individual offset}
\newacronym{CDN}{CDN}{Content Delivery Network}
\newacronym{CN}{CN}{core network}
\newacronym{CP}{CP}{Control Plane}
\newacronym{CPU}{CPU}{central processing unit}
\newacronym{CoMP}{CoMP}{Coordinated Multipoint}
\newacronym{CSR}{CSR}{Cell Site Router}
\newacronym{CQI}{CQI}{Channel Quality Indicator}
\newacronym{C-RAN}{C-RAN}{Cloud RAN}
\newacronym{CS}{CS}{central scheduler}
\newacronym{CSI}{CSI}{channel state information}
\newacronym{CRE}{CRE}{cell range expansion}
\newacronym{D2D}{D2D}{Device-to-Device}
\newacronym{DL}{DL}{Downlink}
\newacronym{DFT}{DFT}{discrete Fourier transform}
\newacronym{DSL}{DSL}{Digital subscriber line}
\newacronym{EARFCN}{EARFCN}{E-UTRA Absolute Radio Frequency Channel Number}
\newacronym{EC}{EC}{European Commission}
\newacronym{eICIC}{eICIC}{enhanced inter-cell interference cancellation}
\newacronym{eMBB}{eMBB}{enhanced Mobile Broadband}
\newacronym{eNodeB}{eNodeB}{Evolved Node B}
\newacronym{EPC}{EPC}{Evolved Packet Core}
\newacronym{EPS}{EPS}{Evolved Packet System}
\newacronym{ETSI}{ETSI}{European Telecommunications Standards Institute}
\newacronym{E-UTRAN}{E-UTRAN}{Evolved Universal Terrestrial Radio Access Network}
\newacronym{FANET}{FANET}{Fly Ad Hoc Network}
\newacronym{FDMA}{FDMA}{frequency division multiple access}
\newacronym{FFT}{FFT}{fast Fourier transform}
\newacronym{FSO}{FSO}{Free-Space Optical Communication}
\newacronym{FTP}{FTP}{File Transfer Protocol}
\newacronym{FU}{FU}{Frame Usage}
\newacronym{GTP}{GTP}{GPRS Tunneling Protocol}
\newacronym{GGSN}{GGSN}{Gateway GPRS Support Node}
\newacronym{GPS}{GPS}{global positioning system}
\newacronym{GRA}{GRA}{Grey relational analysis}
\newacronym{GSM}{GSM}{Global System for Mobile Communications}
\newacronym{GEO}{GEO}{Geosynchronous}
\newacronym{GTP-U}{GTP-U}{GPRS Tunneling Protocol-User Plane}
\newacronym{HAPS}{HAPS}{High Altitude Platform Station(s)}
\newacronym{HDFS}{HDFS}{Hadoop Distributed File System}
\newacronym{HetNet}{HetNet}{Heterogeneous Network}
\newacronym{HiveQL}{HiveQL}{Hive Query language}
\newacronym{HD}{HD}{High Definition}
\newacronym{HEO}{HEO}{High Earth Orbit}
\newacronym{HO}{HO}{handover}
\newacronym{HARQ}{HARQ}{Hybrid automatic repeat request}
\newacronym{HS-DSCH}{HS-DSCH}{High Speed Downlink Shared Channel}
\newacronym{HSS}{HSS}{Home Subscriber Station}
\newacronym{HTS}{HTS}{High Throughput Satellite}
\newacronym{HTTP}{HTTP}{Hypertext Transfer Protocol}
\newacronym{IAB}{IAB}{Integrated Access and Backhaul}
\newacronym{ICIC}{ICIC}{inter-cell interference cancellation}
\newacronym{ICN}{ICN}{information-centric network}
\newacronym{IEEE}{IEEE}{Institute of Electrical and Electronics Engineers}
\newacronym{IMEI}{IMEI}{International Mobile Station Equipment Identity}
\newacronym{IMSI}{IMSI}{International Mobile Subscriber Identity}
\newacronym{IMS}{IMS}{IP Multimedia Subsystem}
\newacronym{IMT-A}{IMT-A}{International Mobile Telecommunications - Advanced}
\newacronym{ITU}{ITU}{International Telecommunication Union}
\newacronym{IP}{IP}{Internet Protocol}
\newacronym{IPsec}{IPsec}{Internet Protocol Security}
\newacronym{IoT}{IoT}{Internet of Things}
\newacronym{ISAC}{ISAC}{Integrated Sensing and Communication}
\newacronym{JSON}{JSON}{JavaScript Object Notation}
\newacronym{KPI}{KPI}{key performance indicator}
\newacronym{LAC}{LAC}{location area code}
\newacronym{LEO}{LEO}{Low Earth Orbit}
\newacronym{LoS}{LoS}{Line-of-Sight}
\newacronym{LPWAN}{LPWAN}{Low Power Wide Area Network}
\newacronym{LTE}{LTE}{Long Term Evolution}
\newacronym{LTE-A}{LTE-A}{Long Term Evolution Advanced}
\newacronym{mmWave}{mmWave}{millimeter wave}
\newacronym{MAC}{MAC}{Medium Access Control}
\newacronym{MADM}{MADM}{Multiple Attribute Decision Making}
\newacronym{MANET}{MANET}{Mobile Ad Hoc Network}
\newacronym{MBH}{MBH}{Mobile Backhaul}
\newacronym{MCS}{MCS}{Modulation Coding Scheme}
\newacronym{MEO}{MEO}{Medium Earth Orbit}
\newacronym{MEW}{MEW}{multiplicative exponent weighting}
\newacronym{MIMO}{MIMO}{multiple-input multiple-output}
\newacronym{ML}{ML}{Machine Learning}
\newacronym{MME}{MME}{Mobility Management Entity}
\newacronym{mMTC}{mMTC}{massive Machine Type Communications}
\newacronym{MMF}{MMF}{max-min fairness}
\newacronym{MMSE}{MMSE}{minimum mean square error}
\newacronym{MPLS}{MPLS}{Multiprotocol Label Switching}
\newacronym{MSISDN}{MSISDN}{Mobile Station International Subscriber Directory Number}
\newacronym{MSP}{MSP}{Mobile Service Provider}
\newacronym{MT}{MT}{Maximum Throughput}
\newacronym{NAS}{NAS}{Non Access Stratum}
\newacronym{NE}{NE}{Nash Equilibrium}
\newacronym{NR}{NR}{New Radio}
\newacronym{NTN}{NTN}{Non-Terrestrial Network}
\newacronym{NFV}{NFV}{Network Functions Virtualization}
\newacronym{NoSQL}{NoSQL}{Not Only SQL}
\newacronym{OAM}{OAM}{Operation, Administration and Management}
\newacronym{OFDM}{OFDM}{orthogonal frequency division multiplexing}
\newacronym{OFDMA}{OFDMA}{orthogonal frequency division multiple access}
\newacronym{ONF}{ONF}{open networking foundation}
\newacronym{ONOS}{ONOS}{Open Network Operating System}
\newacronym{OPEX}{OpEx}{operating expenditure}
\newacronym{OS}{OS}{operating system}
\newacronym{OTT}{OTT}{over-the-top}
\newacronym{PCI}{PCI}{Physical Cell Identity}
\newacronym{PCRF}{PCRF}{Policy and Charging Rules Function}
\newacronym{PDF}{PDF}{Probability Distribution Function}
\newacronym{PDN}{PDN}{packet data network}
\newacronym{PDCP}{PDCP}{Packet Data Convergence Control}
\newacronym{PDSCH}{PDSCH}{Physical Downlink Shared Channel}
\newacronym{PDU}{PDU}{Protocol Data Unit}
\newacronym{PF}{PF}{Proportional Fair}
\newacronym{PGW}{P-GW}{Packet Data Gateway}
\newacronym{PHY}{PHY}{physical layer}
\newacronym{PoC}{PoC}{Proof-of-Concept}
\newacronym{PPP}{PPP}{{P}oisson point process}
\newacronym{PTP}{PTP}{Precision Time Protocol}
\newacronym{QoE}{QoE}{quality-of-experience}
\newacronym{QoS}{QoS}{quality-of-service}
\newacronym{QCI}{QCI}{QoS Class Identifier}
\newacronym{PSC}{PSC}{Primary Scrambling Code}
\newacronym{PSD}{PSD}{power spectral density}
\newacronym{RACH}{RACH}{random access channel}
\newacronym{RAN}{RAN}{Radio Access Network}
\newacronym{RAT}{RAT}{Radio Access Technology}
\newacronym{RB}{RB}{Resource Block}
\newacronym{RE}{RE}{range extension}
\newacronym{RF}{RF}{radio frequency}
\newacronym{RG}{RG}{rate guarantee}
\newacronym{RLC}{RLC}{Radio Link Controller}
\newacronym{RNC}{RNC}{Radio Network Controller}
\newacronym{RR}{RR}{Round Robin}
\newacronym{RRC}{RRC}{Radio Resource Control}
\newacronym{RRH}{RRH}{remote radio head}
\newacronym{RRU}{RRU}{Remote Radio Unit}
\newacronym{RRM}{RRM}{radio resource management}
\newacronym{RSI}{RSI}{RACH Root Sequence Index}
\newacronym{RSS}{RSS}{received signal strength}
\newacronym{RSSI}{RSSI}{received signal strength indicator}
\newacronym{RSRP}{RSRP}{reference signal received power}
\newacronym{RTT}{RTT}{Round Trip Time}
\newacronym{SAC}{SAC}{service area code}
\newacronym{SANET}{SANET}{Sea Ad Hoc Network}
\newacronym{SAW}{SAW}{simple additive weighting}
\newacronym{SC-FDMA}{SC-FDMA}{single carrier frequency division multiple access}
\newacronym{SCN}{SCN}{small cell network}
\newacronym{SCTP}{SCTP}{Stream Control Transmission Protocol}
\newacronym{SDN}{SDN}{Software-Defined Networking}
\newacronym{SDO}{SDO}{Standard Development Organization}
\newacronym{SDMN}{SDMN}{Software Defined Mobile Network}
\newacronym{SDU}{SDU}{Service Data Unit}
\newacronym{SecGW}{SecGW}{Security Gateway}
\newacronym{SGSN}{SGCN}{Serving GPRS Support Node}
\newacronym{SGW}{S-GW}{Serving Gateway}
\newacronym{SHARING}{SHARING}{Self-organized Heterogeneous Advanced RadIo Networks Generation}
\newacronym{SNR}{SNR}{signal-to-noise ratio}
\newacronym{SINR}{SINR}{signal-to-interference-plus-noise ratio}
\newacronym{SISO}{SISO}{single-input single-output}
\newacronym{SSID}{SSID}{Service Set Identification}
\newacronym{ST}{ST}{Standart Multi-User TOPSIS}
\newacronym{STBCs}{STBCs}{space-time block codes}
\newacronym{SyncE}{SyncE}{Synchronous Ethernet}
\newacronym{TB}{TB}{Transport Block}
\newacronym{TBS}{TBS}{Transport Block Size}
\newacronym{TCP}{TCP}{Transport Control Protocol}
\newacronym{TDMA}{TDMA}{Time Division Multiple Access}
\newacronym{TEID}{TEID}{tunnel endpoint identifier}
\newacronym{TOPSIS}{TOPSIS}{Total Order Preference By Similarity to the Ideal Solution}
\newacronym{TTI}{TTI}{transmission time interval}
\newacronym{UAV}{UAV}{Unmanned Aerial Vehicle}
\newacronym{UAV-BS}{UAV-BS}{Unmanned Aerial Vehicles-Base Station}
\newacronym{UARN}{UARN}{UAV-aided relay network}
\newacronym{UDP}{UDP}{User Datagram Protocol}
\newacronym{UE}{UE}{user equipment}
\newacronym{UL}{UL}{Uplink}
\newacronym{UQD}{UQD}{UAV-BS QoS Determination}
\newacronym{UP}{UP}{User Plane}
\newacronym{UMTS}{UMTS}{Universal Mobile Telecommunications Service} 
\newacronym{URLLC}{URLLC}{Ultra-reliable low latency communications}
\newacronym{VANET}{VANET}{Vehicular Ad Hoc Network}
\newacronym{VoIP}{VoIP}{voice over IP}
\newacronym{VPN}{VPN}{virtual private network}
\newacronym{VSAT}{VSAT}{Very Small Aperture Terminal}
\newacronym{W-CDMA}{W-CDMA}{Wideband Code Division Multiple Access}
\newacronym{WiFi}{WiFi}{Wireless Fidelity}
\newacronym{Wi-Fi}{Wi-Fi}{Wireless Fidelity}
\newacronym{WiMAX}{WiMAX}{Worldwide Interoperability for Microwave Access}
\newacronym{WLAN}{WLAN}{Wireless Local Area Network}
\newacronym{WMC}{WMC}{weighted Markov chain}
\newacronym{ZF}{ZF}{zero-forcing}
\newacronym{MNO}{MNO}{Mobile Network Operator}
\newacronym{RES}{RES}{Renewable Energy Sources}
\newacronym{SON}{SON}{Self Organizing Network}
\newacronym{ANR}{ANR}{Automatic Neighbor Relation}
\newacronym{MRO}{MRO}{Mobility Robustness Optimizer}
\newacronym{MLB}{MLB}{Mobility Load Balancing}
\newacronym{CQO}{CQO}{cell quality offset}
\newacronym{BESS}{BESS}{Battery Energy Storage System}
\newacronym{GBS}{GBS}{Ground Base Station}
\newacronym{PV}{PV}{Photovoltaic}
\newacronym{DMC}{DMC}{Disaster Management Center}
\newacronym{SEMA}{SEMA}{Sustainable Energy Management Algorithm}
\newacronym{DEMA}{DEMA}{Disaster Energy Management Algorithm}
\newacronym{EV}{EV}{Electric Vehicle}
\newacronym{MDRU}{MDRU}{Movable and Deployable Resource Unit}
\newacronym{V2G}{V2G}{Vehicle to Grid}
\newacronym{G2V}{G2V}{Grid to Vehicle}
\newacronym{VHetNet}{VHetNet}{Vertical Heterogeneous Network}
\newacronym{IRS}{IRS}{Intelligent Reflective Surfaces}
\newacronym{mMIMO}{mMIMO}{massive multiple-input multiple-output}
\newacronym{WPAN}{WPAN}{Wireless Personal Area Network}
\newacronym{CR}{CR}{Charging Rate}
\newacronym{V2L}{V2L}{Vehicle to Load}
\newacronym{GSP}{GSP}{Grid Selling Price}
\newacronym{GBP}{GBP}{Grid Buying Price}
\newacronym{DSO}{DSO}{Distribution System Operator}
\newacronym{SOC}{SOC}{State of Charge}

\begin{document}
\bstctlcite{IEEEexample:BSTcontrol}

\title{Enhancing  Resiliency  of Integrated Space-Air-Ground-Sea  Networks with Renewable Energies: A Use Case After the \textcolor{black}{2023} Türkiye Earthquake}

\author{Bilal Karaman, Ilhan Basturk, Sezai Taskin, Ferdi Kara, Engin Zeydan, and Halim Yanikomeroglu  
}

\renewcommand{\baselinestretch}{1.1}
\selectfont 

\markboth{IEEE Communications Magazine}%
{Karaman \MakeLowercase{\textit{et al.}}: Enhancing Resiliency of Integrated}
%



\maketitle

\begin{abstract}
Natural disasters can have catastrophic consequences– a poignant example is the series of $7.7$- and $7.6$-magnitude earthquakes that devastated Türkiye on February 6, 2023. To limit the damage, it is essential to maintain the communications infrastructure to ensure individuals impacted by the disaster can receive critical information. The disastrous earthquakes in Türkiye have revealed the importance of considering communications and energy solutions together to build resilient and sustainable infrastructure. Thus, this paper proposes an integrated space-air-ground-sea network architecture that utilizes various communications and energy-enabling technologies. This study aims to contribute to the development of robust and sustainable disaster-response frameworks. In light of the Türkiye earthquakes, two methods for network management are proposed: the first aims to ensure sustainability in the pre-disaster phase and the second aims to maintain communications during the in-disaster phase. In these frameworks, communications technologies such as \ac{HAPS}–which are among the key enablers to unlock the potential of 6G networks–and energy technologies such as \ac{RES}, \glspl{BESS}, and \glspl{EV} have been used as the prominent technologies. By simulating a case study, we demonstrate the performance of a proposed framework for providing network resiliency. The paper concludes with potential challenges and future directions to achieve a disaster-resilient network architecture solution.

\end{abstract}



%
\IEEEpeerreviewmaketitle

\section{Introduction}
%
%
%
%

On February 6, 2023, two powerful earthquakes struck Pazarcık and Elbistan, in Kahramanmaraş, Türkiye – with a magnitude unprecedented in recent history. The earthquakes, with magnitudes of $7.7$ and $7.6$, respectively, caused widespread damage in 11 provinces. The earthquakes have cost more than $48,000$ lives and have destroyed more than $500,000$ buildings, critical communications, and energy infrastructure, resulting in significant financial losses. According to the Turkish Information and Communication Technologies Authority, there are approximately 12 million mobile phone subscribers across the 11 provinces impacted by the earthquake, corresponding to approximately $14$\% of the entire country. The total damage in the telecommunications sector amounts to at least \$$185$ million, according to the Turkish Office of Strategy and Budget \cite{TRdisasterreport}. After the earthquake, satellite communications terminals, mobile \glspl{BS}, emergency communication vehicles, and generators were delivered to the affected region. The power outages in this region were the primary cause of the interruptions to mobile communication and internet services. While mobile \glspl{BS} were sent to the region, limited service was delivered through generators that could merely provide an average of $3$-$4$ hours of energy. As the power outages decreased, communication services began to be delivered for longer periods of time \cite{TRdisasterreport}. 

The series of devastating earthquakes has shown that energy and communication are among the most significant problems during natural disasters. Existing communication infrastructures and network equipment can be damaged in the event of a disaster or can be affected by power failures. As a result, the cellular network or Internet connection is unavailable and the central control authority is limited in its ability to receive timely information regarding the disaster area, further limiting the transmission of critical information \cite{HAZRA202054}. In some instances, high-quality images need to be sent without interruption and with minimal latency for emergency response and life-saving treatments. Some existing energy and communications systems such as \glspl{UAV}, satellite networks, \ac{IoT}, \ac{RES}, \glspl{EV}, etc. are used as solutions to this problem separately \cite{Kwasinski2015, matracia}. However, instead of using these technologies individually, an integrated energy and communication structure should be considered in the development of disaster-resistant communication systems. Furthermore, this integrated network can be enhanced by novel technologies, such as \ac{HAPS} from a communication perspective and \ac{BESS} from an energy perspective. \ac{HAPS} are promising candidates that provide numerous advantages such as coverage adaptability, endurance, rapid deployment, broadband capability, and low cost \cite{di2018haps}. They also offer sustainable solutions for infrastructure damage, energy issues (e.g. power outages), and the restoration of damaged communication infrastructures. 

In the light of above discussion, we propose the utilization of space-air-ground-sea networks with renewable energy to provide resilient and sustainable communications services during disasters. We consider integrated space-based networks (including satellite), air networks (including \ac{HAPS} and \ac{UAV}-based networks), ground networks (including \ac{D2D} communication), and sea-based networks as communication solutions, and \ac{RES} (including wind turbines and solar panels), \ac{BESS}, and \ac{EV} as energy enablers. The disaster management life cycle consists of pre-disaster activities (mitigation and preparedness), response, and post-disaster recovery. Taking into account these phases, for our integrated architecture, two methodologies are also proposed in the pre-disaster phase for sustainability and the in-disaster phase for communication resiliency.

\section{Current Problems of Terrestrial Networks in Post-Earthquake}

\subsection{Infrastructural Damage}
\label{damage}

Ensuring the continuity of communication systems' functionality is of critical importance in disaster situations. \textcolor{black}{\ac{BS} installations are frequently implemented on structures such as buildings and traffic lights, which are susceptible to the impact of disasters.} \textcolor{black}{Depending on the intensity of an earthquake, the backhaul links in \glspl{BS} \textcolor{black}{may} also be damaged.} For example, in the Türkiye earthquake, $25$\% of the existing \glspl{BS} were disabled at time zero, and a \textcolor{black}{significant amount of the fiber infrastructure became unusable}\footnote{\textcolor{black}{Online:} \textcolor{black}{https://m-tod.org/operatorler-deprem-bolgesinde/}, \textcolor{black}{Available: November 2023.}}. \textcolor{black}{Also, while \glspl{BS} on towers suffered relatively less damage, many of the \glspl{BS} located on building tops were either destroyed or damaged. The primary issue was the \glspl{BS} going out of service due to damage to the power grid.} Considering the potential damage to both communication and energy infrastructure in the earthquake, it is crucial to develop integrated energy and communication-based alternative solution scenarios. Note that \ac{HAPS} can play an active role, especially if the backhaul connection is damaged.

\subsection{Energy Problems (\textcolor{black}{Power} Outage, Gas Transportation Problems)}
\label{energyproblems}

\textcolor{black}{During a disaster, power outages can occur due to various reasons such as the collapse of transformers, the toppling of transmission and distribution poles, and the breakage of power lines, etc.} \textcolor{black}{In such circumstances, the restoration of the power grid can be a time-consuming process.} \textcolor{black}{ The endurance time of the generators is $3$-$4$ hours; however, in the Türkiye earthquake, the fuel needed for generators to operate continuously was often not delivered on time due to challenging weather and field conditions \cite{TRdisasterreport}.} \textcolor{black}{As a consequence, the number of disabled \glspl{BS} reached $50$\%, for some operators even on the second and third days after the disaster.} In this regard, \ac{RES}-supported sustainable \ac{GBS}, \ac{RES}-supported Eco Caravan systems, and the use of \glspl{EV} as energy sources can provide significant contributions to the operation of terrestrial networks. In \glspl{VHetNet}, \ac{HAPS} also leverage the advantage of having a huge payload capacity. This enables them to utilize solar energy and batteries to sustain uninterrupted operation for months or even years \cite{kurt}. Hence, it will become possible to prevent problems associated with power outages during disasters.

\subsection{Restoring Communication}
\label{restoring}

Restoring damaged communications infrastructure after a disaster or due to power supply problems is a vital problem that often takes a long time. The time taken to restore operations was also too long in Türkiye earthquake. The existing mobile \glspl{BS} were not enough, and mobile \glspl{BS} were even brought in from different countries, and this process took days. As can also be seen from the Cloudflare statistics, the full capacity could already be reached again in one month after the earthquake \footnote{\textcolor{black}{Online:} https://blog.cloudflare.com/q1-2023-internet-disruption-summary/, \textcolor{black}{Available: November 2023.}}. To this end, thanks to their strong \ac{LoS} communications capability and a combination of \ac{RF} and \ac{FSO} for backhauling, \ac{HAPS} nodes are able to provide communications services without the need for additional ground infrastructure. \textcolor{black}{However, HAPS could have challenges, e.g; in terms of communication and sensitivity to environmental changes. Nevertheless, HAPS can be supported by various technologies and its effectiveness in reaching remote or densely populated disaster areas can be improved.  By using technologies such as \ac{mMIMO} and \ac{IRS}, \ac{HAPS} can leverage precise beamforming and steering techniques.  This enables \ac{HAPS} to provide wireless connectivity over robust \ac{LoS} connections, covering a radius of up to several hundred kilometers \cite{gunes}. \ac{HAPS} can also use the S-band to react more sensitively to changes in the environment.}

\section{{\textcolor{black}{Proposed}} Space-Air-Ground-Sea Integrated Network Architecture}

\begin{figure*}[htp!]
\centering
\includegraphics[width=\linewidth]{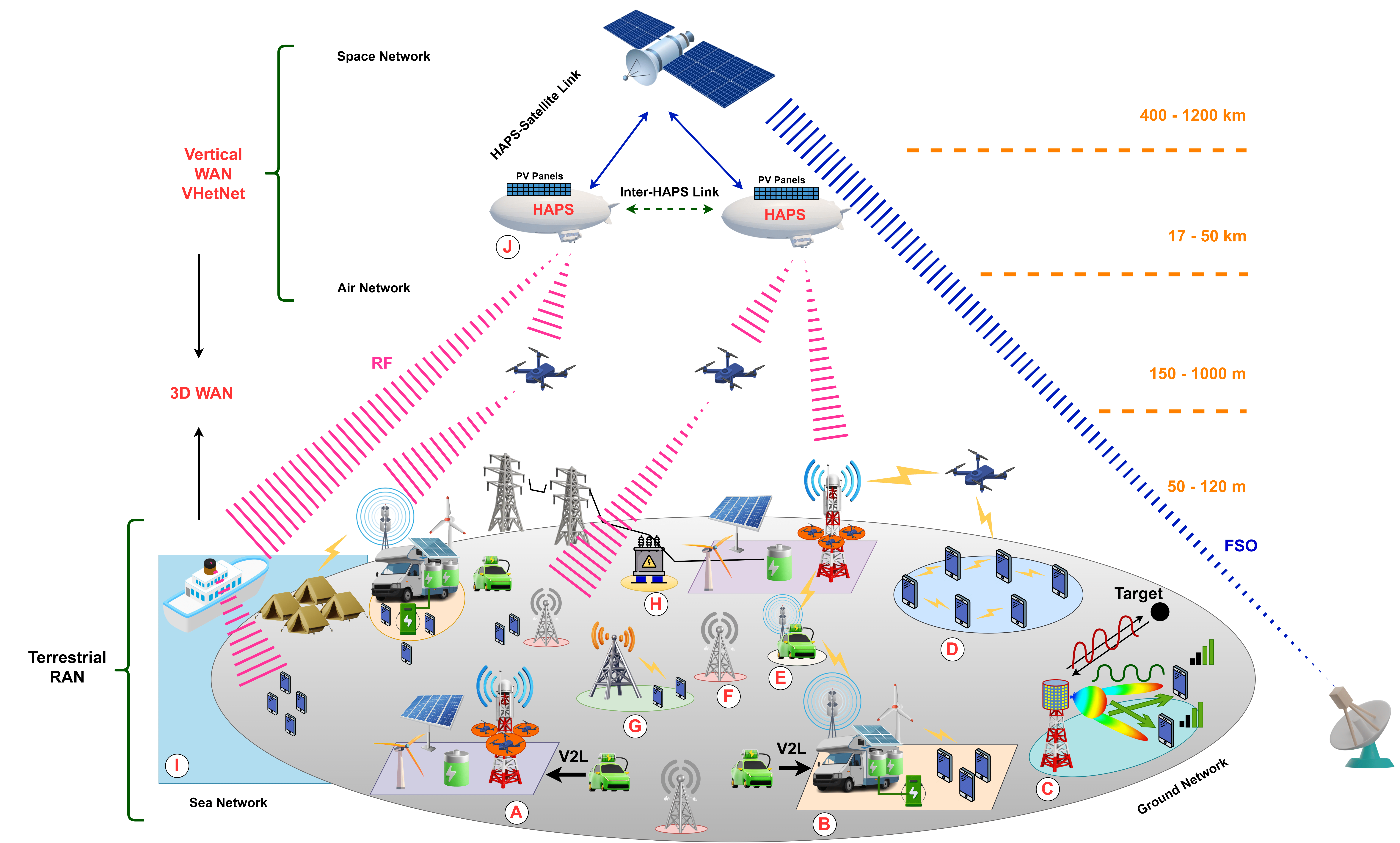}
\caption{Proposed integrated space-air-ground-sea network topology with energy and communication solutions.}
\label{disaster_v1}
\vspace{-0.5cm}
\end{figure*}

In this section, we propose a \ac{HAPS}-supported architecture, which can be used as both \ac{RAN} and backhaul. \ac{HAPS} mainly refers to airborne systems stationed at high altitudes, typically in the stratosphere, above traditional aviation airspace and can be in the form of drones, balloons, or airships. Fig. \ref{disaster_v1} shows the \textcolor{black}{proposed integrated network architecture that combines many energy technologies like  \ac{RES}, \glspl{BESS} and \glspl{EV} and  many communication technologies like \ac{LEO} satellite systems, HAPS, UAVs, \ac{ISAC} etc at different layers, namely space, air, and terrestrial}. In Fig. \ref{disaster_v1} the area in gray represents the disaster-struck zone. The shaded areas are as follows: (i) \textit{A} represents \ac{RES} supported sustainable \ac{GBS}, which includes a \ac{BESS}. (ii) \textit{B} represents the \ac{RES}-supported Eco Caravan system as using \ac{MDRU}. (iii) In \textit{C}, an \ac{ISAC} is indicated. (iv)  \textit{D} represents the \ac{D2D} communication. (v) As shown in \textit{E}, \glspl{EV} can play an important role in disasters as a source of energy. In addition, \glspl{EV} can be used as mobile relays. (vi) \textit{F} represents the failed \glspl{BS} and (vii) \textit{G} represents the active \glspl{BS} in the disaster. (viii) \textit{H} represents seismic isolator-based solutions to improve the resistance of critical power system components such as transformers to seismic vibrations. (ix) \textit{I} represents the sea-based networks. (x) \textit{J} indicates a \ac{HAPS}-based communication system. \textcolor{black}{Some of the prominent communication and energy technology enablers used in the proposed architecture for the disaster scenarios are explained in the following subsections}.

\subsection{Communication Technology Enablers}

\textbf{Space-Based Networks} can operate independently of terrestrial networks while providing a reliable and resilient communications infrastructure and emergency response support in remote or hard-to-reach areas. Satellites can play a crucial role in disaster response communication planning, particularly in the context of earthquakes, due to their wide coverage, resilience, and ability to provide connectivity in remote or damaged areas. Satellite networks can be used to establish temporary communication infrastructure in disaster-prone areas and communication channels between emergency responders, local communities, and other stakeholders.  Rescue teams, government agencies, and relief organizations can use satellite connectivity to share vital information, coordinate resources, and efficiently manage rescue and relief operations.  Satellite networks can also be used to share critical information such as maps, images, and video between emergency responders and command centers \cite{rajput2020impact} and warn and disseminate important information to local communities.   This information aids in identifying critical infrastructure damage, assessing the extent of the disaster, and guiding resource allocation and deployment.

\textbf{Air-Based Networks} contains novel technologies like HAPS and UAVs. \ac{HAPS} can play a critical role in improving the resilience of integrated space-air-ground-sea networks during and after natural disasters such as earthquakes as shown in Fig. \ref{disaster_v1} \textit{area J}. Some of the advantages of \ac{HAPS} are: (i) Because of its surface area, it has the potential to generate almost its own energy with solar panels. (ii) A single \ac{HAPS} can be an alternative to all damaged terrestrial \glspl{BS} because it can provide a large coverage area. Here, a single \ac{HAPS} can be perceived as a multi-sector (directional) \ac{BS}.
(iii) For almost all outdoor users, a direct \ac{LoS} can be established that can compensate for distance-related pathloss or scattering losses. (iv) Compared to satellites (e.g., Starlink), no additional intermediate devices (e.g., relays) are required. Moreover, satellites (e.g., \ac{LEO}) can establish a \ac{LoS} of maximum 6 minutes to any ground station. For this reason, thousands of satellites were launched to create a constellation, but \ac{HAPS} can provide stable and continuous service because they do not make orbital movements. (v) They are very easy and quick to run in terms of logistics (especially useful in the first days after the earthquake, when it becomes impossible to reach the earthquake area by land). (vi) It is an important alternative not only on the \ac{RAN} side, but also for the backhaul. A HAPS can easily forward all backhaul traffic to an unaffected area via links such as \ac{FSO}/THz communication with a remote \ac{HAPS} or ground station. During earthquakes, fiber optic lines can be damaged in the earthquake area, so some \glspl{BS} may be out of service. (vii) \textcolor{black}{It is not physically affected by weather events.}

\glspl{UAV} can also be a valuable tool for disaster communication planning and warning. It can be used to build temporary communications infrastructure in disaster-prone areas providing a versatile platform for gathering and disseminating critical information in disaster-prone areas and supports emergency response efforts. \glspl{UAV} equipped with cameras or other sensors can provide real-time information about the extent of damage, the location of survivors, and other critical information that can be used to support emergency response efforts. 



\textbf{Ground-Based Networks}   consist of various components, including cellular networks, terrestrial Internet (that relies on physical cables, such as fiber optics or copper wires, to transmit data packets over long distances), and mobile ad hoc networks. Unlike air-based networks, the topology of the ground-based networks is relatively stable and less mobile. In the event of a natural disaster, such as a hurricane, earthquake, or flood, the ground-based communication infrastructure can be severely damaged or destroyed as pointed out in Section \ref{damage}. This can make it difficult for emergency responders and survivors to communicate with each other. One way to overcome this challenge is to use \ac{D2D} communication as shown in Fig. \ref{disaster_v1} \textit{shaded area D}. \ac{D2D} communication provides a decentralized and resilient communication solution in earthquake scenarios. It is especially useful in scenarios where the use of the power grid is inaccessible and the use of batteries for power is not possible.  \ac{ISAC} can also play a critical role in improving the resilience of integrated space-air-ground-sea networks with renewable energy, as shown in Fig. \ref{disaster_v1} \textit{shaded area C} \cite{liu2022integrated}.  \textcolor{black}{Firstly, the extent of the damage caused by an earthquake can be quickly estimated with ISAC. Sensors can be used to monitor the integrity of buildings and communication systems can be used to transmit this information in real time.  Secondly, ISAC can be used to find survivors trapped in the rubble. Sensors can be used to detect movement and signs of life, while communication systems relay this information to rescue teams.  Third, ISAC can provide real-time information about the situation on the ground, including the location of survivors, the extent of damage and the status of rescue efforts.  Finally, ISAC can facilitate communication between response teams so that they can coordinate their efforts and share information in real-time. This can ensure that resources are allocated effectively and relief efforts are optimized. }

\textbf{Sea-Based Networks} assumed critical significance during Türkiye earthquake, where traditional communication infrastructure in coastal provinces collapsed. In this network topology, various elements such as ships and unmanned surface vehicles can facilitate communication by leveraging \glspl{UAV}, \ac{HAPS}, and satellite technologies based on their respective locations as shown in Fig. \ref{disaster_v1} \textit{shaded area I}. \textcolor{black}{They can complement the proposed architecture by improving coverage and resiliency by offering diverse paths and reducing the risk of network congestion and outages. They can also facilitate interoperability between different types of networks (e.g., naval vessels, aircraft, and ground-based command \& control centers) via standard or customizable protocols enabling seamless communication between different services and agencies.}

\subsection{Energy Technology Enablers}

The importance of energy in the design and operation of future networks cannot be overstated, as global energy consumption is estimated to increase by $20$\% by 2030, driven by the growing number of subscribers, traffic demand, connected devices, and services driven by information and communication technologies \cite{kement}. Another critical factor is ensuring the resilience of power systems to maintain uninterrupted communications in the event of disasters, both during and after they occur. The term \enquote{power system resiliency} refers to the ability of an electrical power system to mitigate the magnitude and duration of disruptive events. Some of the possible solutions in this context are: (i) The introduction of smart grids to replace centralized traditional grid infrastructures opens up the possibility of incorporating distributed generation units and microgrids based on renewable energies on a significant scale, and thus has the potential to play a decisive role in disaster scenarios. (ii) Ensuring a reliable power supply through the use of energy storage systems, such as \ac{BESS}, is of paramount importance. (iii) The use of \ac{EV} batteries in \ac{V2L} applications is proving to be a viable approach. (iv) Implementing solutions based on seismic isolators is a promising approach to improve the seismic resilience of critical power system components to strong shaking.

\textcolor{black}{The United Nations aims to achieve net-zero CO$_2$ emissions by 2050 to mitigate the impacts of climate change \cite{BM}. In order to achieve this goal, \ac{RES} are a growing area of energy systems to enable decarbonization and sustainability policies.} With the high penetration rate of \ac{RES} such as wind and solar energy and the participation of \ac{RES}-supported microgrids and \ac{RES}-supported \glspl{BS}, the energy system becomes more resilient to disasters. In addition, it will be possible to operate terrestrial communications systems for a much longer period of time during disasters. \textcolor{black}{When considering air networks,} \textcolor{black}{t}he large surface area of \ac{HAPS}, the lack of cooling systems required for communications and energy components, and the potential to meet nearly all power needs through the integration of \ac{PV} panels makes the use of \ac{RES} in air-based networks feasible.

\subsection{\textcolor{black}{Integrated Approach}}

\textcolor{black}{In this section, we provide some examples of the joint use of energy and communication enablers. In Fig. \ref{disaster_v1} \textit{shaded area A}, it is proposed to install \ac{GBS} towers in certain locations, especially in densely populated areas, and to support these \ac{GBS} towers with \glspl{BESS}. In this context, the \ac{GBS} is aimed to produce its own energy with \ac{PV} panels and wind turbines. The \ac{GBS} can autonomously charge the \ac{BESS} from the \ac{RES} and in case of a power outage, it can utilize the stored energy from the \ac{BESS}. Charging stations will also be available on the \ac{GBS} towers, allowing the charging of multiple \glspl{UAV}. \glspl{UAV} can be directed to desired areas under the control of the Disaster Management Center, and \glspl{UAV} with low battery levels can also be diverted to charging stations for recharging. \ac{UAV} charging stations can be used not only in emergency situations, but also to support areas with high communication needs, agricultural or meteorological activities, observations, \ac{IoT} applications, and various purposes such as communication with different sensor nodes.} \textcolor{black}{Note that \glspl{MDRU} play an important role in the communication system and their energy requirements are mostly covered by diesel generators. \glspl{MDRU} can be transformed into the Eco Caravan supported by \ac{RES}, which is shown in Fig. \ref{disaster_v1} \textit{shaded area B}. The aim is to have a portable vehicle equipped with a \ac{BESS}, solar and wind energy. Thanks to \glspl{EV}, the energy requirements of both \ac{GBS} and \glspl{MDRU} can be met. In addition, thanks to \ac{GBS} and \glspl{MDRU} with \ac{BESS}, wireless power transmission technology for user devices can be evaluated in the event of a disaster.}

In disasters, terrestrial communications can suffer severe damage. In Fig. \ref{disaster_v1} \textit{area J}, a hybrid architecture combining satellite and terrestrial infrastructures with \ac{HAPS} can be deployed. Integration of \ac{FSO} and \ac{RF} technologies can be used to create integrated uplink networks connecting space, air, sea, and ground systems. \glspl{UAV}, especially \ac{HAPS}, can serve as relay stations in this setup to increase reliability. Alternatively, a standalone solution relying solely on the satellite and \ac{HAPS} segment can be used when terrestrial infrastructure is not available. \ac{HAPS} can be easily set up, quickly forwarded to disaster areas, repaired, and easily replaced for reconfiguration. Moreover, owing to their substantial payload capacity, \ac{HAPS} can be outfitted with \ac{PV} modules, enabling the provision of wireless service for extended durations, spanning several months. The wide coverage area provided by \ac{HAPS} makes this network more cost-effective compared to deploying numerous terrestrial networks. This is how a ubiquitous and disaster-resistant communication architecture can be implemented via \ac{HAPS}.

\vspace{-0.3cm}
\section{\textcolor{black}{Two Use Cases for Türkiye} \textcolor{black}{Earthquake}}

\textcolor{black}{In this section, we propose two hybrid energy and communications methodologies tailored to the earthquake in Türkiye using the architecture described above with communication and energy components. The first methodology aims at ensuring sustainable integrated HAPS-Terrestrial networks in the pre-disaster situation. The second methodology aims to prioritize communication and resiliency in the event of an earthquake. The proposed methodologies are explained in the following subsections.}

\subsection{\textcolor{black}{Methodology 1}}

\begin{figure}[htp!]
\centering
\includegraphics[width=\linewidth]{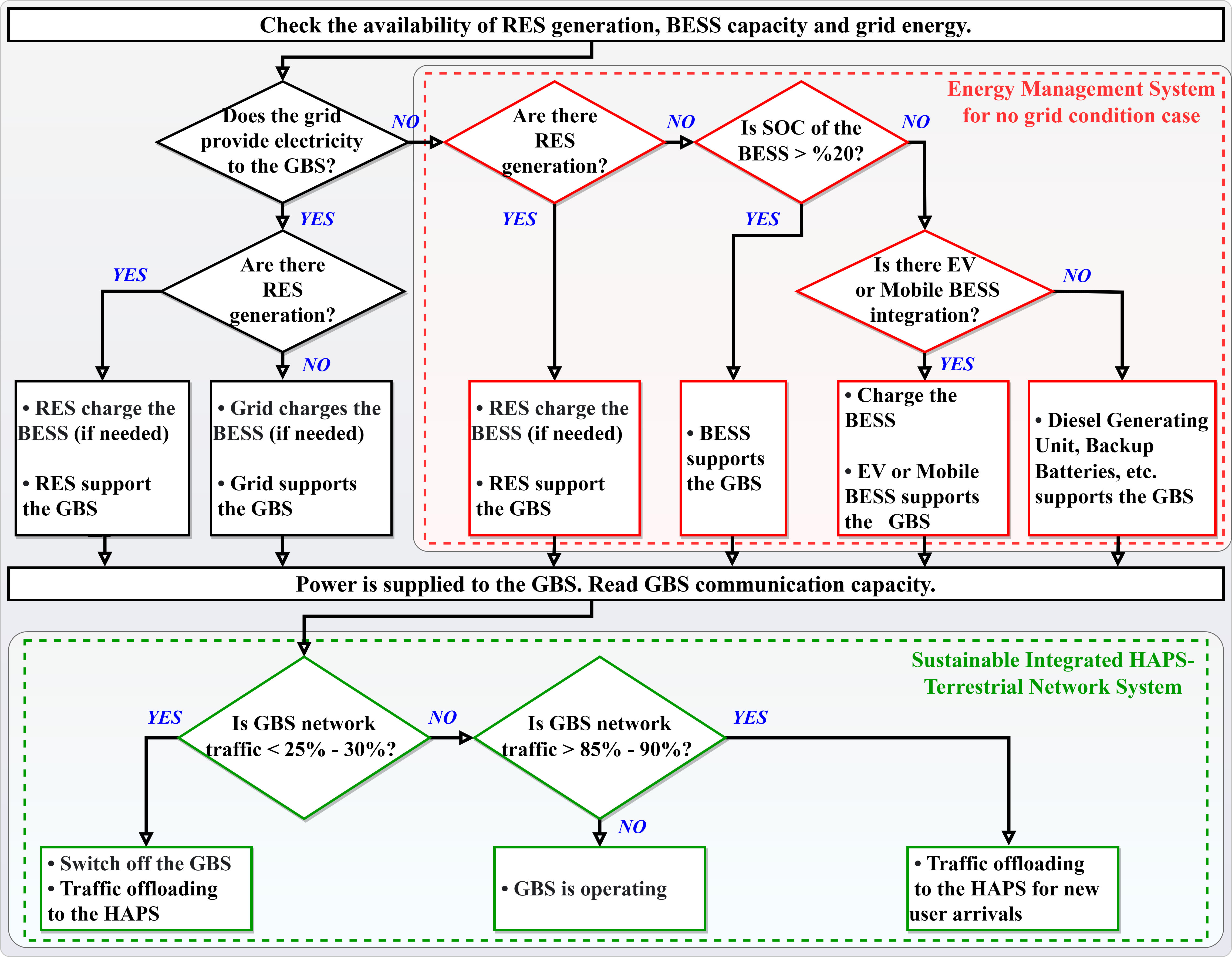}
\caption{\textcolor{black}{Methodology 1 for pre-disaster.}} 
\label{sustainable_gbs_ema}
\vspace{-0.3cm}
\end{figure}

\textcolor{black}{The proposed hybrid methodology for pre-disaster depicted in Fig. \ref{sustainable_gbs_ema} consists of a zero-touch energy and network management system. This methodology is primarily about ensuring sustainability. The hybrid methodology consists of energy management with \ac{RES} and \ac{BESS} and the communication management system with \ac{HAPS} to ensure sustainability in \ac{GBS}.  In \cite{kement}, it was shown that dynamic traffic management and energy savings in dense urban areas can be achieved by switching off terrestrial \glspl{BS} with low network traffic and routing traffic to \ac{HAPS}. Therefore, in our hybrid methodology, while \ac{GBS} has low network traffic, the traffic is offloaded to \ac{HAPS} by switching off the \ac{GBS}. Besides, it is proposed to offload the traffic for new user arrivals to \ac{HAPS} when \ac{GBS} is approaching congestion status. This also eliminates the need of network densification to meet dynamic and unprecedented data traffic in the highly-dense areas that is conventionally handled by adding additional permanent/temporary \ac{GBS}. In this regard, the proposed methodology promises sustainability in pre-disaster scenarios. According to detailed analysis in \cite{kement}, the proposed HAPS-integrated methodology is also economically viable in terms of capital and operational expenditure.}

\subsection{\textcolor{black}{Methodology 2}}

\begin{figure}[htp!]
\centering
\includegraphics[width=\linewidth]{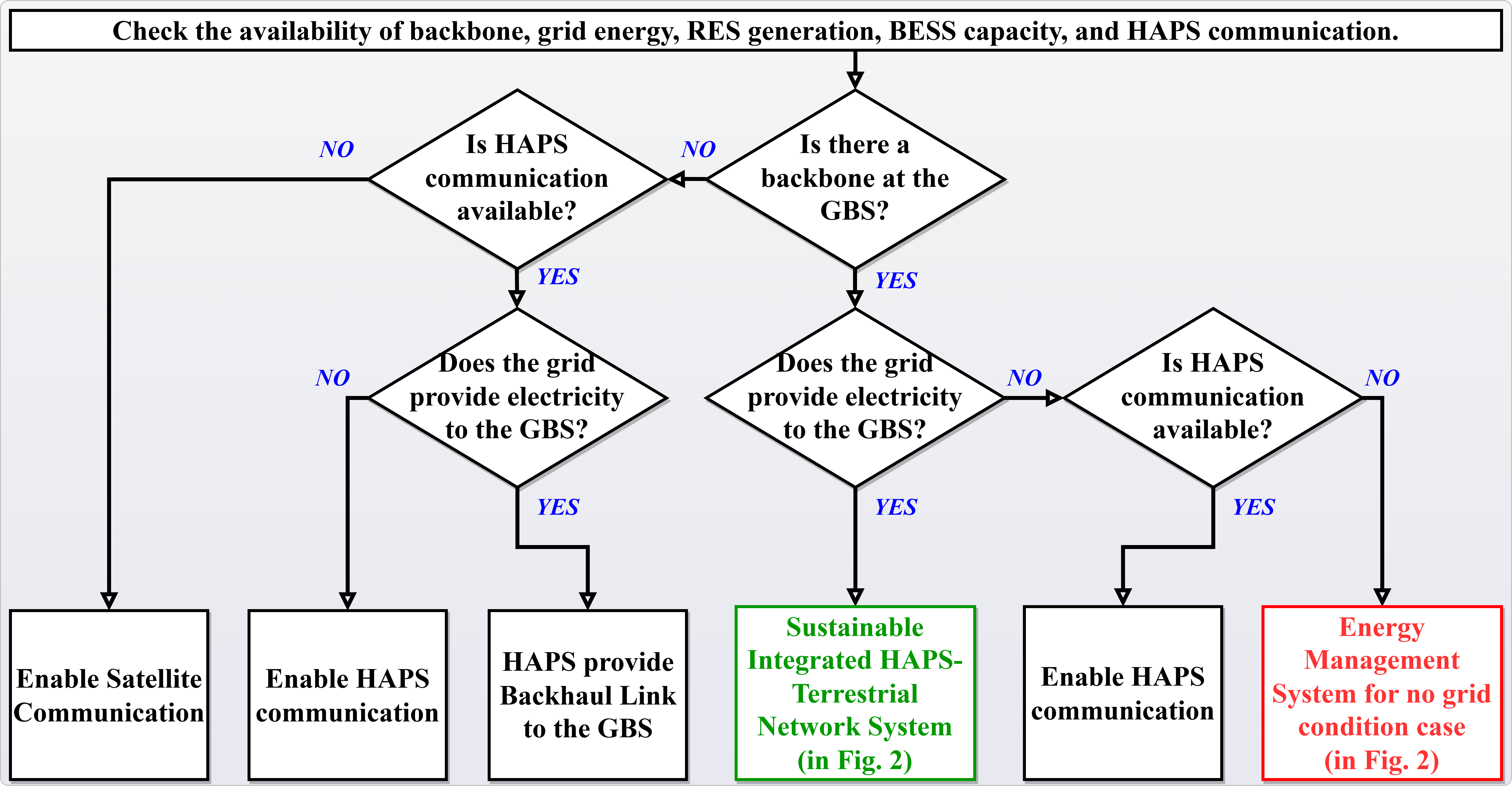}
\caption{\textcolor{black}{Methodology 2 for in-disaster.}}
\label{disaster_ema}
\vspace{-0.3cm}
\end{figure}

\textcolor{black}{The proposed hybrid methodology for the disaster case shown in Fig. \ref{disaster_ema}, aims to prioritize communication in the event of an earthquake. This methodology aims to ensure communication via the \ac{HAPS} and to protect the batteries of the \ac{GBS} if the \ac{GBS} is deactivated due to a power failure. In cases where communication with \ac{HAPS} is lost or the service is not available, \ac{GBS} aims to primarily meet the energy requirements with \ac{BESS} considering the \ac{SOC}. In scenarios where the \ac{BESS} is not charged, proposing a solution involves directing volunteer \glspl{EV} or mobile \glspl{BESS} to the \ac{GBS} for \ac{BESS} charging. Another alternative for energy needs is the use of other supporting resources such as backup batteries on \ac{GBS}, diesel generator units, etc. In the scenario where the \ac{GBS} is disabled due to an interruption in the fiber optic line (backbone), the aim is to provide a backhaul connection with the \ac{HAPS} if the \ac{GBS} has energy over the grid. In the event of a power failure in the \ac{GBS}, all communication is handled via HAPS. In cases where communication with \ac{HAPS} is lost or the service is not available, satellite communication can be considered as an alternative.}

\subsection{\textcolor{black}{Simulation Results for the Türkiye Earthquake Case Study}}

\begin{figure}[]
\centering
\subfloat[Mobile handheld devices that operate in S-band. ]{\includegraphics[scale=0.5]{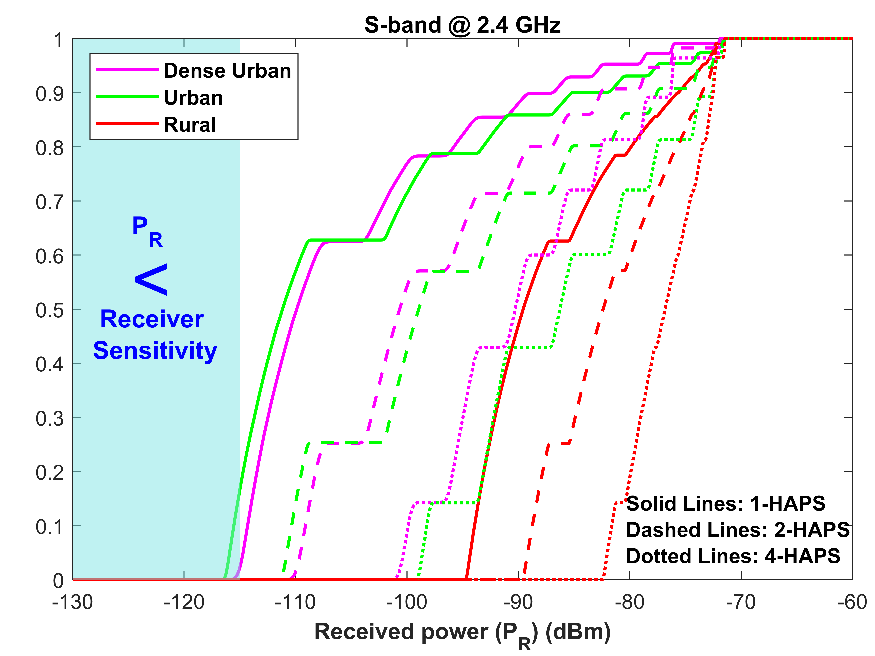}
\label{cdf1}} \\
\subfloat[Very small ($\approx 60$cm) aperture terminal users operate in Ka-band.]{\includegraphics[scale=0.5]{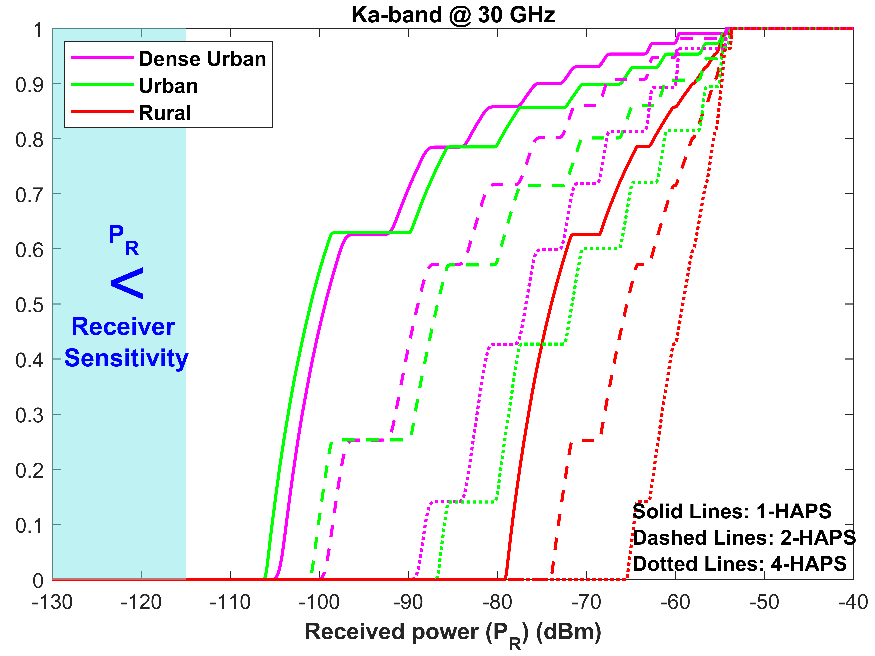}
\label{cdf22}}
\caption{CDF of received power at the ground terminals for a HAPS-RAN communication in the whole affected region.}
\label{vs_SNR}
\vspace{-0.7cm}
\end{figure}

\textcolor{black}{In order} to demonstrate the \ac{RAN} capabilities of \ac{HAPS} networks \textcolor{black}{in Fig. \ref{disaster_ema}} - \textcolor{black}{Methodology 2}, we show in Fig. \ref{vs_SNR} computer simulation results for the \ac{CDF} of receiver power at ground users. As stated in \cite{3GPP_UE}, the \ac{RAN} \textcolor{black}{connection} between a \ac{HAPS} (or \ac{HAPS} constellation) and the users on the ground can be established in both S- and Ka-bands. In S-band, \textcolor{black}{mobile} handheld devices are supported, while in Ka-band Very Small Aperture Terminals (VSAT) (i.e., $60$ cm circular polarisation) are required, \textcolor{black}{which} can be permanently installed or mounted on mobile platforms (e.g., vehicles, UAVs, etc.). \textcolor{black}{During} the earthquakes in Türkiye, the total affected area (11 provinces) is about $115,000$ km$^2$ with a total of 12 million mobile subscribers \cite{TRdisasterreport}. Therefore, for the S-band simulations (Fig. \ref{vs_SNR} (a)), all $12$ million mobile subscribers are randomly distributed in the area (for simplicity) and served by a single HAPS, $2$-HAPS or $4$-HAPS. In contrast, in the Ka-band simulations (Fig. \ref{vs_SNR} (b)),  \textcolor{black}{we assume that $1$ million VSAT subscribers} are randomly distributed in the same area and served by the same amount of HAPS network. In all simulations, the parameters of HAPS (e.g., transmit power, antenna gain, etc.) are used based on \cite{3GPP_NR}. Air-to-ground channel characteristics (e.g., \ac{LoS} probability, path loss, shadowing, atmospheric loss, etc.) and \textcolor{black}{the} receiver parameters (e.g., antenna gain, sensitivity, etc.) are also taken from \cite{3GPP_UE}. \textcolor{black}{From} the extensive simulations, we can conclude that even a single HAPS could restore all communications \textcolor{black}{in the region} - \textcolor{black}{and thus replace} all failed ground BSs -\textcolor{black}{as} none of the users receive a signal lower than the receiver sensitivity. Nevertheless, by increasing the number of HAPS, we were able to improve \ac{QoS} thanks to the $10$-$20$ dB increase in received SNR by each HAPS. As expected, \textcolor{black}{the} performance deteriorates in the dense urban scenarios due to lower \ac{LoS} probability and higher losses (e.g., shadowing, clutter losses, etc.) \cite{3GPP_UE}. On the other hand, in Fig. \ref{vs_SNR} (b), we can observe an average $20$ dB higher received power than in Fig. \ref{vs_SNR} (a) in all scenarios. This can be explained as follows. Although the path loss is \textcolor{black}{greater} in Ka-band, the antenna gain of the VSAT receiver can compensate for this and \textcolor{black}{provide} a higher \ac{QoS} \cite{3GPP_NR}. \textcolor{black}{In summary, our simulation results show that even in the scenario in which all terrestrial \glspl{BS} are deactivated, the communication needs in the earthquake-damaged region can be met with \ac{HAPS}.}

 \begin{table*}[htp!]
 \begin{center}
\caption{Comparison of Communication Enablers for different KPIs in earthquake scenarios \textcolor{black}{with proposed hybrid methodologies}.}
\label{tab:Compare}
\scalebox{0.8} {
\begin{tabular}{|c|c|c|c|c|c|c|c|c|c|} 
 \hline
\backslashbox{\textbf{KPIs}} {\textbf{Technology}}& Satellite  &  HAPS & UAV  & Terrestrial & D2D & \textcolor{black}{Methodology 1} & \textcolor{black}{Methodology 2}         \\ 
 \hline\hline
Latency & High & Medium & Low  & Low & Low  & \textcolor{black}{Medium}  & \textcolor{black}{Medium}    \\ 
 \hline
 \rowcolor{lightgray}
Bandwidth & High & High & Medium & Medium & Low & \textcolor{black}{High}   & \textcolor{black}{High}   \\
 \hline
  Resiliency & High & High & Medium & Low  & Medium & \textcolor{black}{Medium}    & \textcolor{black}{High}   \\
 \hline
 \rowcolor{lightgray}
   Scalability & High & High &Medium & Medium & Medium &  \textcolor{black}{High}   & \textcolor{black}{High}   \\
    \hline
   Security  & High & High & Medium & High &  Medium & \textcolor{black}{Medium}   & \textcolor{black}{Medium}   \\
 \hline
 \rowcolor{lightgray}
   Interoperability  & Medium & Medium & Medium & High & Low & \textcolor{black}{Medium}   & \textcolor{black}{Medium} \\
 \hline
   Cost-efficiency & Medium & High  & Medium  & Medium & Low & \textcolor{black}{High} & \textcolor{black}{High}   \\
    \hline
 \rowcolor{lightgray}
   Energy efficiency & Medium & High & Medium &Low & High & \textcolor{black}{High} & \textcolor{black}{Medium}  \\ 
 \hline
\end{tabular}}
\end{center}
\label{enablers_status}
\vspace{-0.5cm}
\end{table*}

\textcolor{black}{Our proposed methodologies combine terrestrial and \ac{HAPS} technologies, but their objectives are different depending on the  scenario considered. To compare our integrated hybrid methodologies with} the primary communication enablers, Table \ref{tab:Compare} \textcolor{black}{provides} \textcolor{black}{comparisons} \textcolor{black}{in terms of} various \glspl{KPI}. \textcolor{black}{According to the \glspl{KPI} in Table 1, both \textit{Methodology 1} and \textit{Methodology 2} perform medium in terms of interoperability, latency and security but performs high in terms of bandwidth and scalability. Compared to \textit{Methodology 2}, the proposed \textit{Methodology 1} is better in terms of energy efficiency. However, \textit{Methodology 2} provides better communication resilience.}

\section{Challenges and Future Research Directions}

\textcolor{black}{The proposed architecture includes the use of different technologies, protocols, and strategies to ensure seamless communication and energy support in disaster scenarios. The challenges of coordinating the above elements include ensuring seamless interoperability, overcoming communication latency issues, and managing limited resources in a dynamic and often chaotic disaster environment. How future networks can be managed to ensure seamless connectivity, interoperability, and efficient resource utilization must be considered as a future research direction. Note that there are some proposals for solving these problems in the field of integrated networks \cite{Darwish, farajzadeh2023self}, which could be used to find solutions. To achieve a disaster-resilient network architecture, some of the future solutions are as follows:} \textbf{(i) \ac{AI} and blockchain networks:} Solutions that rely on \ac{AI}-based techniques can be used to solve various disaster communication planning problems both in pre-disaster, in-disaster and post-disaster scenarios. In addition, \ac{AI} together with blockchain frameworks can improve remote data protection and secure energy trading. \textbf{(ii) Cloud native deployments:} To improve network scalability and flexibility of networks in disaster zones, cloud-native deployments of network services can be used in space-air-ground-sea networks. This can also reduce development cycles, capital, and operational expenditures while enabling platforms to be open. Container orchestration tools like Kubernetes can help with cloud-native deployments. \textbf{(iii) Energy harvesting:} If the user's devices are able to use radio, solar or vibration energy from the environment, energy consumption constraints can be significantly reduced.

\section{Conclusions}

In this paper, an integrated space-air-ground-sea network architecture is used, which uses various communication and energy-enabling technologies, to improve resiliency and sustainability in disaster scenarios. The integrated communication network presented includes satellite, \ac{HAPS}, \ac{UAV}, and \ac{D2D} technologies. In addition, the integration of various \glspl{RES} (e.g., wind turbines and solar panels) together with innovative energy enablers (e.g., \ac{BESS} and \glspl{EV}) is proposed to ensure sustainable energy management in disaster situations. Specifically targeting the recent earthquakes in Türkiye, two network management methodologies are put forth. The performance of the proposed framework is validated through the simulation of a case study demonstrating the strength of \ac{HAPS} communication in disaster areas. Future research and practical implementation in this direction holds great potential for the development of more resilient and sustainable disaster management strategies.

\bibliographystyle{IEEEtran}
\bibliography{bibliography}

\vspace{-1.8cm}

\begin{IEEEbiographynophoto}
{Bilal Karaman} (bilal.karaman@cbu.edu.tr) is a Research Assistant at Manisa Celal Bayar University, Türkiye. He is also a Visiting Researcher at Carleton University, Ottawa, Ontario, Canada.
\end{IEEEbiographynophoto}
\vspace{-1.8cm}
\begin{IEEEbiographynophoto}
{Ilhan Basturk} [M] (ilhan.basturk@cbu.edu.tr) is an Associate Professor at Manisa Celal Bayar University, Türkiye.
\end{IEEEbiographynophoto}
\vspace{-1.8cm}
\begin{IEEEbiographynophoto}
{Sezai Taskin} (sezai.taskin@cbu.edu.tr) is a Professor at Manisa Celal Bayar University, Türkiye.
\end{IEEEbiographynophoto}
\vspace{-1.8cm}
\begin{IEEEbiographynophoto}
{Ferdi Kara} [SM] (f.kara@ieee.org) is an Assistant Professor at Zonguldak Bulent Ecevit University, Türkiye. He is also a Postdoctoral Fellow at KTH Royal Institute of Technology, Stockholm, Sweden.     
\end{IEEEbiographynophoto}
\vspace{-1.8cm}
\begin{IEEEbiographynophoto}{Engin Zeydan} [SM] (engin.zeydan@cttc.cat)  is a Senior Researcher at Centre Tecnològic de Telecomunicacions de Catalunya (CTTC), Spain.
\end{IEEEbiographynophoto}
\vspace{-1.8cm}
\begin{IEEEbiographynophoto}{Halim Yanikomeroglu} [F] (halim@sce.carleton.ca)  is a Chancellor's Professor at Carleton University, Ottawa, ON, CA.
\end{IEEEbiographynophoto}




\end{document}